\documentstyle[11pt,newpasp,twoside,epsf]{article}
\def\edcomment#1{\iffalse\marginpar{\raggedright\sl#1\/}\else\relax\fi}
\def\etal{{\it et al.}~}
\def\halpha{{H$_{\alpha}$}~}
\def\hbeta{{H$_{\beta}$}~}

\reversemarginpar

\begin{document}
\title{Magellanic Cloud Planetary Nebulae: A Fresh Look at the 
Relations between Nebular and Stellar Evolution}
 \author{L. Stanghellini $^{1,2}$, R. A. Shaw $^1$, J. C. Blades $^1$,
B. Balick $^3$}
\affil{$^1$ Space Telescope Science Institute, 3700 San Martin Drive, Baltimore
(MD) 21218.}
\affil{$^2$ Affiliated with the Astrophysics Division, Space Science
Department, ESA}
\affil{$^3$ Department of Astronomy, University of Washington}

\begin{abstract}
Studies of the relationship between planetary nebula morphology and the 
evolution of the central stars has long suffered from uncertainties in 
distance determinations, and from the bias of interstellar absorption, 
that are typical for Galactic PNe. We will be able to eliminate the 
distance errors and be assured of the sample homogeneity by studying 
Large Magellanic Cloud (LMC) PNe with images from the 
{\it Hubble Space Telescope}. 
In this talk we present the first observations in our new sample. 
The data consist of broad-band images and medium dispersion, slit-less 
spectra obtained with STIS, and are of excellent quality. 
Indeed, these data show great promise for subsequent analysis, which will 
centered on the relationship between nebular morphology and stellar and 
nebular evolution.  While the most intensive analysis of the sample must 
await the completion of the survey, the data obtained so far show that 
we will learn a lot along the way.

\end{abstract}

\section{Why study Magellanic Cloud Planetary Nebulae?}

Planetary Nebulae (PNe) and their central stars (CSs) have been used for 
decades as probes to the final phases of low- and intermediate-mass stars. 
The nebular gas expelled at the end of the Asymptotic Giant Branch (AGB), 
and sculpted by radiation and a CS fast wind, acquires a shape that 
depends upon the combination of the distribution of gas and dust in the 
circumstellar shell, and upon the local ISM 
environment. It it thus important to study nebulae
and stars together to obtain a more complete understanding of 
the final phases of stellar evolution. 

Since the pioneering work by Greig (1972), PN morphology has been associated 
to stellar population and, by inference, to the CS mass (see Stanghellini 
1995 for a review); the latest finding on the correlations between nebular
morphology and stellar evolution/population (e.g., Manchado, this volume)
certainly show that there is a strong trend that connects large masses stars
and asymmetric PNe.  Nonetheless, to date the (sometime very) convincing 
trends that link the evolution of stars and nebulae have been criticized 
for the biases and uncertainties inherent in the distance scale for Galactic 
PNe.  On these grounds, we decided to pursue the investigation in the
Large Magellanic Cloud, where uncertain distances are not an issue. 
A study of this type is presently only possible with the capability 
of the {\it Hubble Space Telescope}.  
In this short communication we present the aims our project,
some limited analysis of SMP-27, and a roadmap to future developments.

\section{The LMC Strategy}

The Large Magellanic Cloud (LMC) PNe have a typical size of $\sim$half an 
arcsecond. For this reason, their morphology (and size) make them suitable
targets for the HST.  As of this writing, the only published HST data on 
Magellanic Cloud PNe are monochromatic images obtained by Dopita (Dopita 
\etal 1996) using WFPC-1, and Blades \etal (1992) using FOC (see also 
Stanghellini \etal 1999), on a grand total of 30 PNe.  Both sets suffer 
from spherical aberration of the HST primary mirror prior to the first 
servicing mission.  
A third set of LMC PNe is available in the HST Data Archive (PI = Dopita), 
which includes 14 objects observed in monochromatic lines with WFPC2.  
Clearly, to perform an LMC PN study of the type of that reported by 
Stanghellini \etal (1993) for Galactic PNe, 
there is the need for a lager database. 
For this reason we pursued our slit-less spectroscopic study of 50 LMC PNe.  
We chose to obtain slitless spectra in order to determine the PN shapes in 
a range of different ionization
levels. In particular, we aim at obtaining {\it monochromatic}
images in the major nebular lines such \hbeta, [O III], \halpha, [N II],
[O I], He I and [S II]. 

\section{STIS slitless observation of SMP-27}

Our program was approved as a HST Cycle 8 snapshot. 
When deciding on our target list, we choose PNe within a large span in the 
\hbeta and [O III] fluxes. Images of the PNe 
were obtained with the Space Telescope Imaging Spectrograph (STIS) in 
white (50CCD) light, and with the medium-dispersion gratings G430M and G750M, 
which allowed us to obtain the morphologies in the traditional lines ([O III], 
\halpha, and [N II], as well as the others mentioned above.
The \halpha image of SMP-27 is shown in Figure 1. The slitless spectrograms
provides both the spectral and spatial information, thus in this image we 
see the emission nebula and the material surrounding it. 
The planetary nebula SMP-27 is very clearly quadrupolar, and shows this
type of morphology in all prominent spectral lines. The ring segment and
the blob of outer material that we were able to observe around this
nebula are suggestive of a previous ejection event from the PN 
progenitor, or, possibly, a remnant of the prior AGB fast wind.  The 
absence of an obvious, nearby source of ionizing radiation, and the 
decline in the ionization level (as measured by the ratio of the [O III] 
and \halpha images) makes the association of the distant arc and blob 
with the CS fairly secure.

\begin{figure}
\plotone{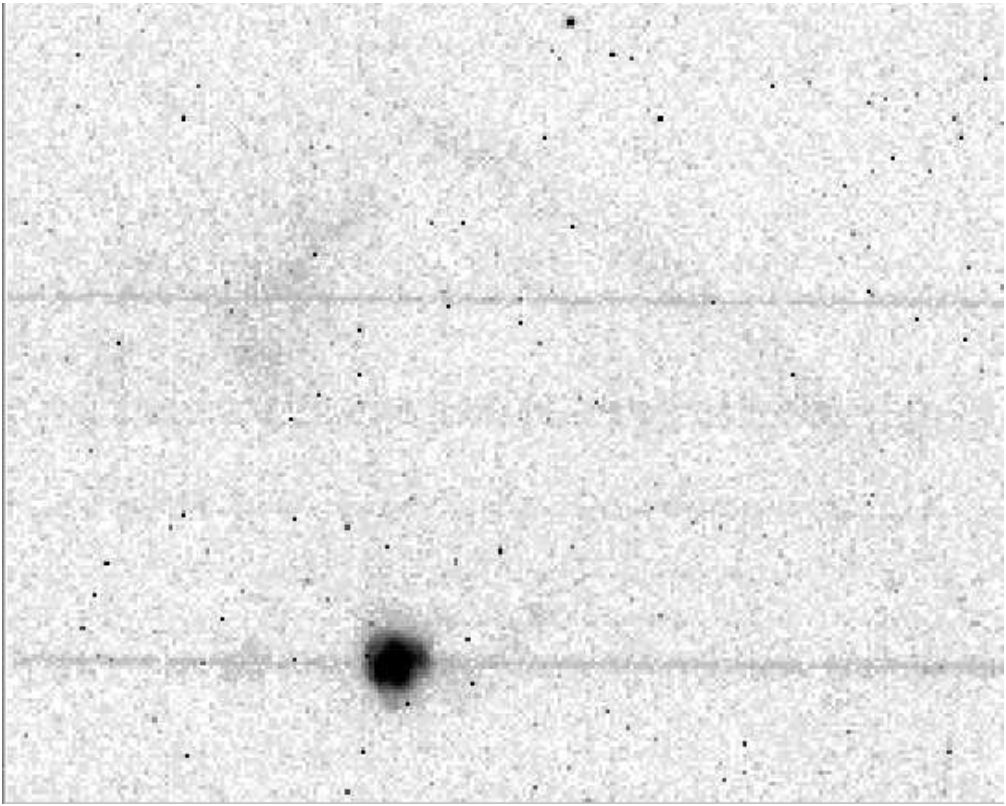}
\caption{\halpha image of LMC SMP-27 (North up, East left).
Note the quadrupolar shape, and the
outer material possibly ejected at previous stellar stages.}
\end{figure}

A quick analysis shows that SMP-27 is a high excitation PN 
of approximately 0.85 arcsec of diameter (encircling 95 \% of the
nebular light), corresponding to 0.106 pc. The separation between
the main nebular body and the north-west arc is about 1.53 pc, while the arc
itself is 2 pc long. There is also a fainter blob of material 14.6 deg east
of north.
%The nebular fluxes measured are the following: F(\halpha)=8.5e-14;
%F(\hbeta)=3.21e-14; F($\lambda$ 5007 [O III])=2.12e-13, all in cgs units.
The nebula is also detected in the follwoing lines: $\lambda$ 6678 He I,
$\lambda$ 6584 [N II], $\lambda$ 6548 [N II], $\lambda$ 4959 [O III].
The calculated reddening for this PN is zero. The star has a hot 
spectrum, more prominent in the blue than in the red, which is consistent 
with the relatively high nebular ionization. The estimated magnitude of the 
CS is V$\approx$19.6. A preliminary estimate of the luminosity and temperature 
yields log L$=$2.7 and log T$_{\rm eff}=4.5$ by using the hydrogen line 
intensities. As this PN appears to be optically thin, the derived L \& T 
are probably lower limits to the actual luminosity and temperature.

\section{Future work}

At the time of writing, this project just getting underway.
But the exceptional data quality prompted us to present a
preview of the results.  
The overall analysis of the morphological sample will be 
done once the bulk of the observations are complete. 
Morphology, size, expansion age, and central star physics will be studied 
for the homogeneous dataset. 
Ground based spectroscopy will be used, together with the available data 
in the literature, to probe the relationship between morphology and stellar
evolution, including the yield of the PN progenitors.

\end{document}